# Effect of Wind Intermittency on the Electric Grid: Mitigating the Risk of Energy Deficits

Sam O. George, *Member, IEEE*, H. Bola George, Ph.D. and Scott V. Nguyen, Ph.D.

*Abstract*--Successful implementation of California's Renewable Portfolio Standard (RPS) mandating 33 percent renewable energy generation by 2020 requires inclusion of a robust strategy to mitigate increased risk of energy deficits (blackouts) due to short time-scale (sub 1 hour) intermittencies in renewable energy sources. Of these RPS sources, wind energy has the fastest growth rate—over 25% year-over-year. If these growth trends continue, wind energy could make up 15 percent of California's energy portfolio by 2016 (wRPS15). However, the hour-to-hour variations in wind energy (speed) will create large hourly energy deficits that require installation of other, more predictable, compensation generation capacity and infrastructure. Compensating for the energy deficits of wRPS15 could potentially cost tens of billions in additional dollar-expenditure for fossil and / or nuclear generation capacity. There is a real possibility that carbon dioxide and other greenhouse gas (GHG) emission reductions will miss the California Assembly Bill 32 (CA AB 32) target by a wide margin once the wRPS15 compensation system is in place. This work presents a set of analytics tools that show the impact of short-term intermittencies to help policy makers understand and plan for wRPS15 integration. What are the right policy choices for RPS that include wind energy?

*Index Terms*--California, renewable energy, risk analysis, systems engineering, wind power generation.

## I. INTRODUCTION

THE government of California (CA) has set the goal of generating 33% of total energy consumed within the state from renewable power sources, so-called Renewable Portfolio Standard (RPS33) [1][2], by the year 2020. RPS33 is a key component of California's Global Warming Solutions Act of 2006 (Assembly Bill 32) that seeks to reduce $CO_2$ and other GHG to 1990 levels by 2020. The portfolio of renewables includes wind, geothermal, biomass, solar thermal and solar photovoltaic. As of April 2009, about 3.5% of California's energy is derived from wind [3]. Of the completed and operational projects within CA, wind accounts for 3856 GWh/year and represents 26% of the RPS total, second only to geothermal accounting for 57% (8353 GWh/year) of the RPS total. Wind constitutes a larger fraction of the approved contracts currently under construction: 43% (6822 GWh/year) of the expected capacity through 2012. The large investment in harnessing wind energy underscores its prominence within the class of renewables and partly serves as the basis for this work. It should be noted that these analyses can be easily extended to solar-based renewables without any loss of specificity.

The primary motivation for this work is to assist policy / decision makers in formulating RPS33 implementation policies with the least systemic risk. The current / prevailing planning tools are based on annualized computations (so-called macro analyses). For renewable energy, such macro analyses mask important hourly (short time-scales; 1 to 3 hour) intermittencies in the data. These variations represent significant systemic risk that, among other consequences, increase the implementation cost of RPS33 beyond current projections. For example, computations that involve averaging over long (annual) time scales might suggest to a policy maker that energy portfolios with the highest proportions of energy from wind sources would be most effective in reducing $CO_2$ and other GHG emissions. However, hourly analytics show the opposite. These short term intermittencies increase the risk of energy deficits as the fraction of RPS in the grid grows. To mitigate the risk of power outages, other, more reliable, power plants (such as fossil-based fuel) need to be maintained in sub-optimal standby mode (so-called "spinning reserve") to rapidly compensate for energy deficits due to intermittencies of renewables. Given the energy deficit risks that increase as the fraction of wind increases, the question remains: at what critical threshold does reliance on wind energy break down? In the scenario that California cannot import additional reserve energy, this critical threshold is about 5% RPS, assuming the state's energy reserve capacity is 5 GWh (see section entitled "Implications").[1]

To our knowledge, two published reports exist that appear to contradict our key finding that higher fraction of energy from wind increases the systemic risk of energy deficits. A cursory glance at the study on why wind power works in Denmark suggests successful implementation of RPS20 from wind [4]. A closer look, as the author discusses, reveals a grid

Financial support for this work is provided by GridByte, Inc., *Energy Policy Analytics* Practice.

S. O. George and H. Bola George, Ph.D. are with GridByte, Inc., 65 Enterprise, Aliso Viejo, CA 92656 USA (e-mail: info@gridbyte.com).

S. V. Nguyen, Ph.D. is with Shell Projects and Technology, Innovation and R&D Division, Houston, TX 77002 USA.

[1] Important: The grid maintains a very tight frequency and power factor balance between demand and generation. The California grid has spinning reserves and other reactive (capacitive) storage assets that offer 5 to 10% (about 2.5 GWh to 5 GWh) reserve capacity to compensate for short imbalances between generation and demand within seconds (or minutes). The systemic risks addressed in this work are due to sustained hour-long imbalances (deficits and surpluses) between generation and load demand.

that is intimately linked to larger grids of neighboring countries (Germany and Sweden) with energy portfolios that rely on coal and large hydro. It is safe to say that without energy from its neighbors, Denmark would experience severe energy deficits (or protracted blackouts) due to the intermittencies or unavailability of wind.

A study commissioned by the state of Minnesota (MN) in 2006 on integration of wind concluded that generation at a penetration level of 20% could be supported if transmission lines were available [5]. The MN analysis employs averaging wind speeds over long (monthly) time scales; thus, the presence of large dips (random periods during which wind speeds are too low) in the wind profiles are largely absent in their work. The analytics we perform actually utilizes some raw data from the MN report, yet our conclusions are different: energy portfolios with high wind penetration levels are subject to a higher risk of energy deficits and blackouts. Compensating for this systemic risk exacerbates the emission of $CO_2$ and other GHGs into the atmosphere as coal- (and other fossil-) fired power plants must be maintained in spinning reserve.[2]

## II. WIND ENERGY ANALYTICS

### A. Definition of Systemic Risk as used in this work

Formally, risk is the product of two components: (The probability of an Energy Deficit) × (The Impact of Energy Deficits). This work focuses on quantifying the probability (and magnitude) of energy deficits. The full treatment of both will be presented in our next report.

We define systemic risk two ways: As California's energy needs grow, the state looks increasingly to renewables for new capacity. In this token, we look to renewables whenever older "dirty" plants are decommissioned. The first systemic risk is that the state does not have adequate reserve energy capacity beyond the normal 5 to 10% (about 2.5 to 5.0 GWh) [7]. Thus, as the fraction of energy from wind and solar increases beyond 5%, there is no significant capacity to mitigate the risk of energy deficits. The second systemic risk is directly due to the nature of the intermittencies in wind energy production. As we see in Fig. 1 (or Fig. 2), wind speed (or energy) shows random abrupt changes from hour-to-hour. If the energy deficits are not compensated quickly, the grid may experience many blackouts. The state currently has about 381 MW of nameplate peaker capacity [8] that may be used as fast compensators. However, the large random abrupt changes in wind energy outputs require a special class of generation equipment. While peaker generators are designed to provide 20 to 30% of nameplate power during peak demand, their cycling operation are not designed for frequent hour-to-hour changes as would be common in a wRPS15 scenario.

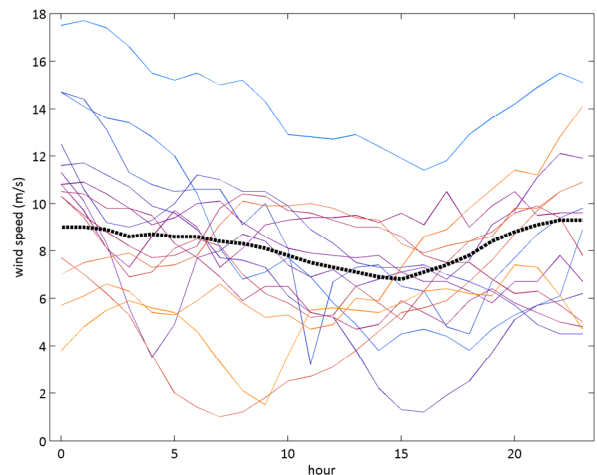

Fig. 1: Diurnal wind speeds (m/s) measured at 80 m for alternating days for Tower 71 in MN during April 2005. The dashed black line represents the mean diurnal wind speed over the entire month.

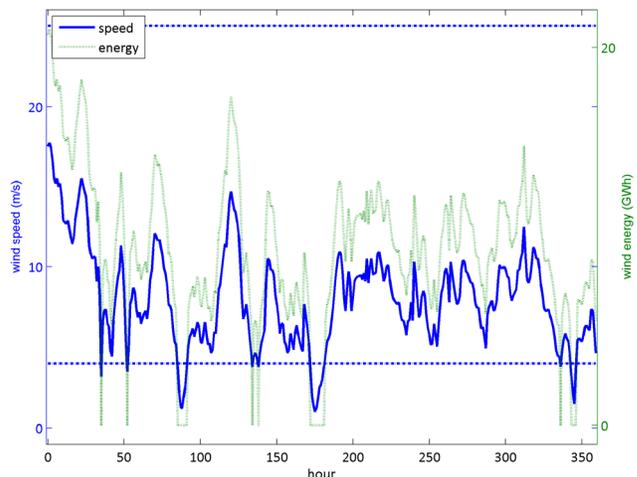

Fig. 2: Wind speed and associated energy profiles for a scenario containing 15 contiguous days. No energy is produced when wind speed either falls below 4 m/s or exceeds 25 m/s. In this case, no energy is produced in 26 (7.22%) out of 360 hours.

### B. California as a case study—24 hour consumption profiles

California is a good case model given its aggressive goal of producing 33% of its energy from renewable sources by the year 2020. Of all the states, energy consumption in CA accounts for 14.79% of all energy produced in the US.[3] To date, CA has the largest nameplate capacity of energy from renewable sources. Fig. 3 shows the total energy production in CA from all sources (left axis, blue squares) and from wind sources (right axis, green circles) by year [10].

To provide a forecast scenario of the penetration-percentage of wind energy, we present an exponential least-squares fitting of historic data projected through the year 2020 in Fig. 4. The year-to-year growth rate of total US consumption is about 2%. Combined with knowledge of the percentage generated from wind, we estimated the yearly fraction of energy generated from wind will increase according to the trajectory shown in Fig. 4--our projections show that CA will achieve wRPS15 by the year 2016 ± 5 months. In Fig. 4, the red curve was obtained from a least squares fit of the data using an exponential function; the r-squared coefficient is 0.837. The parameters of the best fit equation are shown in the legend.

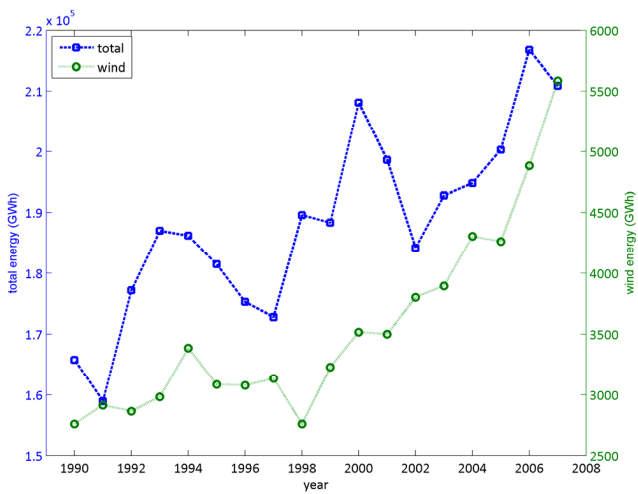

Fig. 3: Annual production of electricity (GWh) in CA from wind (green circles) and all sources (blue squares) by year.

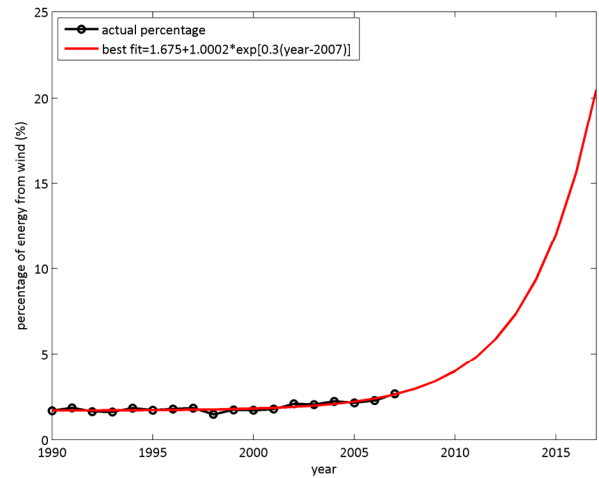

Fig. 4: Percentage of energy generated from wind relative to the total by year. The red curve was obtained from a least squares fit of the data using an exponential function; the r-squared coefficient is 0.837. The parameters of the best fit equation are shown in the legend.

To answer the central question of whether CA can achieve wRPS15, we expand the analysis space by studying a data set of daily energy consumption profiles. Irrespective of percentage from renewables, California's total daily energy consumption profile resembles Fig. 5; daily energy consumption in CA during summer 1999 from residential (blue circles), commercial (red squares), and industrial and agricultural (green triangles) sectors [11]. The solid black trace is the sum from all sources—demand is highest between the hours of 11 AM and 6 PM. We show the 1999 data in Fig. 5 because it is complete and similar in magnitude to 2008. Seasonal and weather-related anomalies do not change the overall profile significantly. The profile may shift up or down to match seasonal variations without loss of accuracy.

## C. Data sources & assumptions

One of the biggest obstacles we have faced is collection of accurate and consistent data. The data used in this work are assembled from several credible sources—and within these, there are some inconsistencies. We have made some assumptions to compensate for incomplete or unavailable data:

1. There are no transmission constraints on the California grid; i.e., all energy generated at wind farms can be dispatched to all consumption markets. This is a hypothetical construct that produces best case estimates.
2. Hour-by-hour energy generation is equal to hour-by-hour energy consumption. From other research we have conducted, we do not forecast significant utility-scale energy storage assets by 2020.
3. Hour-by-hour wind speed data are taken from the state of MN during the month of April 2005 [12]. We assume that the diurnal fluctuations are representative and similar to those in CA. The MN data were more readily available to us.

4. In our models, 33% RPS is applied on an hour-to-hour basis.
5. The annual variation in wind speed was obtained from sites in Tehachapi, CA [13].[4] We assume that the annual trend in the Tehachapi area is representative of profiles across the state over the course of the year.
6. Daily and annual changes in the demand curve in CA are small, i.e., the shape of the demand curve is the same throughout the year. In practice, we expect higher energy demand during summer and winter months; we expect that the reduced demand during spring or fall months would shift the entire demand curve downwards without significant changes in the shape. On a weekly basis, we expect that higher demand in the residential sector will be offset by lower demand in the commercial sector over weekends; we assume that demand in the agricultural sector

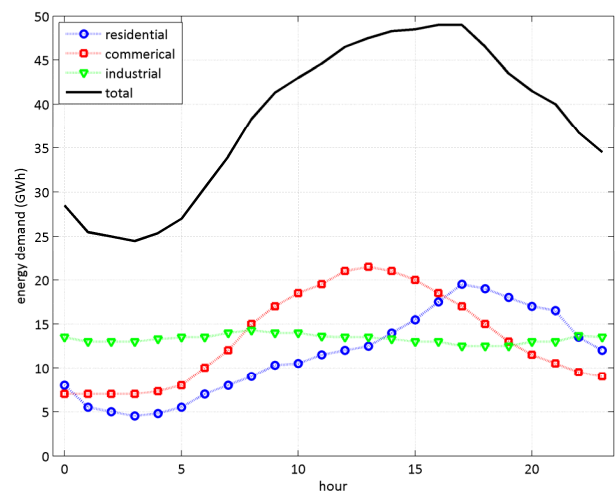

Fig. 5: Daily energy consumption in CA during summer 1999 from residential (blue circles), commercial (red squares), and industrial and agricultural (green triangles) sectors. The solid black trace is the sum from all sources—demand is highest between the hours of 11 AM and 6 PM.

---

[4] The Tehachapi area produces a significant fraction of California's wind energy output.

does not change throughout the week.

We assume near-perfect correlation between wind speed and generated energy. This is the best case. In reality, there are significant wind shear exponents and non-zero lag time between wind speed and availability of *dispatchable* energy. These, combined with the slow response of wind turbines, add to the intermittencies of wind generation. In converting from speed to energy, we have followed the convention that an average daily demand of 1 GWh corresponds to an average daily production of 1 GWh; this allows a scaling factor for conversion from average wind speed (m/s) to energy (GWh).

### D. Future energy deficits and the wRPS15 system design & implementation problem

The gross impact of energy deficits becomes apparent only when the hour-to-hour daily wind energy data are modeled with no statistical averaging. A priori, wind energy is not as "controllable" as fossil sources. Thus it is not possible for wind generation to match any demand profile on an hour-to-hour basis. As we show later in this section, wRPS15 system design that relies on statistical averaging leads policy makers to believe, erroneously, that wind energy can be thought of in the same macro terms as fossil-based sources. Macro energy planning means: 1 GWh of wind energy is exchangeable with 1 GWh of fossil-based energy.

Current macro energy generation design requires a high degree of stability from hour-to-hour. The 5 to 10% reserve capacity in California's grid infrastructure is not designed for the fast hour-long responses necessitated as the fraction of wind energy increases. This scenario necessitates the inclusion of an array of fast-response (1 to 3 hours) (possibly peaker) generation capacity to compensate for hour-to-hour wind intermittencies.

To illustrate our methodology, let us work through a simplified modeling algorithm that produces the composite waveforms in Fig. 6. Again, we present a subset of data to aid visualization. Our modeling includes data points dating back to 1990.

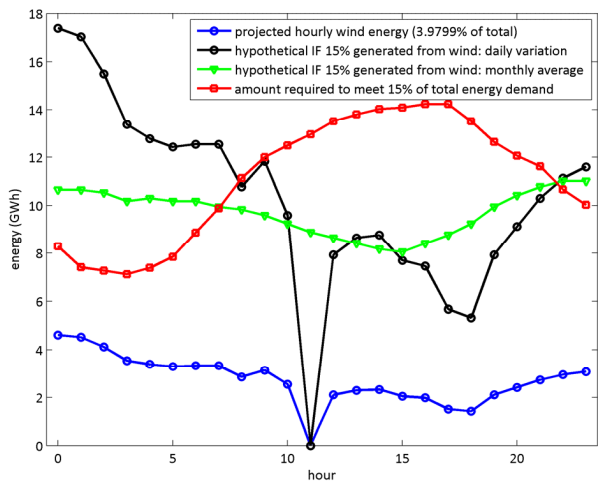

Fig. 6: Daily profiles of energy generated from wind (blue, black and green) in CA on an hourly basis. Calculation of the blue trace assumes a wind penetration level of 4% (obtained by extrapolation of the fit shown in Fig. 4) during the month of March 2010. If a wind penetration level of 15% is assumed, the black and green traces are obtained from actual hourly data within a 24-hour period and averaging by hour over multiple days during a given month, respectively. The red trace represents the amount of energy needed to meet 15% of daily energy demand.

[Simplified Algorithm]

Step 1: The algorithm reads all available wind energy (or speed) data to create 24-hour wind generation basis profiles. The basis profile may be actual data or data from future projection. In Fig. 6, the basis model illustrated is for a day in March 2010 based on the forecast equation in Fig. 4 that shows wind energy penetration at 4% by mid 2010. The 24-hour basis profile is the blue trace in Fig. 6.

The algorithm applies a range of physical constraints to ensure that the energy production is realistic from hour to hour adjusted for seasonal variations. For example, wind speed must be at least 4 m/s and cannot exceed 25 m/s. On the upper end, the turbine blades must be feathered to prevent mechanical damage.

Step 2: Scale the 24-hour wind generation basis profile by the wRPS percentage. In Fig. 6, the basis profile is scaled by 15% to produce the black trace. The reader immediately sees that this wind energy generation profile exhibits large variation during the day. These wind profiles, as shown in Fig. 1 (or Fig. 2), are generally not predictable from day to day. The lack of predictability from day to day (lack of day-ahead forecasting) is the subject of many current wind reports. Our position is different. Even with forecasting, wind speeds cannot be estimated with any degree of high precision. Mitigating systemic risk due to wind intermittency requires an array of fast hour-by-hour compensation equipment.

Step 3: Superimpose energy consumption demand profile from data in Fig. 5. There are literally thousands of daily demand profiles—the red trace in Fig. 6 is one representative day in March 1999 scaled by 15% to match the wind generation basis profile percentage.

The energy production design problem is complicated because the time-varying deficits are very large.

For example, Fig. 6 shows a deficit of about 2.91 GWh at 10AM and 12.95 GWh at 11 AM. These large time-varying deficits show the systemic risk in stark detail. For perspective, the average utility-scale generator in CA has a nameplate rating under 0.25 GW.[5] Thus, in one best-case scenario, CA would need between twenty and sixty utility-scale fast response generating plants to compensate for these random wind energy deficits. Additional insights are discussed in detail in the "Implications" section.

To show how statistical averaging leads to an erroneous macro solution, we superimpose the energy production profile (green trace) comprising the daily hourly statistical averages for the entire month of March 2010 in Fig. 6. This contrast is important because the macro energy planning methods utilize statistical averaging over months (or years) to illustrate relatively stable wind energy capacity. However, as shown by the error bars in Fig. 7 (or fluctuations in Fig. 2), the hour to hour wind energy production probability distributions are large. Compared with the energy demand curve, there are strong probabilities that the energy compensation system would need to produce 20 to 80% of hourly demand.

---

[5] 64% of generation plants in California produce energy in the range 0 to 25 MW [14].

Fig. 7 presents a near complete probability picture of the hour to hour wind energy production compared with the 15% energy demand curve (solid black curve). Fig. 7 is one of the best statistical illustration tools to visualize the inherent probability distribution of wind energy. The hourly distributions are generated from the 15 sample profiles presented in Fig. 1. Fig. 8 shows the surpluses and deficits for each of the 15 daily profiles. As indicated by outliers or whiskers in Fig. 7, the systemic risk due to low wind-energy production is prevalent. Potentially large surplus energy produced (e.g., from 1AM to 3AM) also represents a systemic risk; there is no systemic advantage because of the absence of utility scale storage.

If the state is to fulfill 100% of its energy demand 100% of the time, there are only two likely solutions to eliminate the systemic risk due to energy deficits from wind energy. These are presented in the "Solutions" section.

### E. Implications of the inherent risk of wRPS15

Extending Fig. 7, Fig. 9 shows an energy deficit view superimposed on the RPS15 demand curve. In this picture, we see that the probability deficit distributions of hourly wind energy variations are large. With energy deficits as large as eighty percent of the wRPS15 demand, system risk implications are numerous. We discuss some of these implications in this section.

Compensating for deficits requires planners to abandon macro design principles based on averaging over months or years. As we show above, macro planning methods based on averaging expose the state to random and frequent energy deficits (or surpluses). We list some of the numerous cost implications below.

1. From our analytics, compensating for random probability energy production inherent in wRPS15 requires planning for wind energy far in excess of the macro exchange equation that suggests 1 GWh of wind energy may replace 1 GWh of fossil-based energy. To make wind energy reliable, the total nameplate capacity must exceed the demand profile by a possible factor of 10. And this fundamentally assumes that the wind is blowing. The overdesigned nameplate provisioning represents a capital cost that greatly exceeds available budgets for wRPS implementation / compensation.

2. As reported by others [16], wind energy may simply be unavailable for 20 to 30% of the time. In this work, the fifteen-day scenario shown in Fig. 2 shows no-wind conditions for 26 (7.22%) out of 360 hours. These conditions occur when wind speed falls below 4 m/s or exceeds the maximum usable value of 25 m/s. The only way to compensate for lack of wind energy is to have additional baseload and fast response capacity available in the system. For instance, China's aggressive construction of power plants to harness energy from wind has been accompanied by the "less publicized" construction of coal-fired plants to mitigate blackouts associated with wind intermittency. This is the so-called dirty secret of wind energy production recently reported in the Wall Street Journal [17] as follows: "China's ambition to create "green cities" powered by huge wind farms comes with a dirty little secret: dozens of new coal-fired power plants need to be installed as well." In short, the state needs to maintain an array of fast fossil-fired plants (FFP) to compensate for wRPS unreliability.

3. Energy generation is not something that can be readily turned on and off. The fastest non-RPS power plants require time to ramp up to full capacity. Most energy compensation plants need to be run in standby mode so they can quickly respond to surges in demand. It takes one to two hours [16][18] to maximize the output from a plant in standby mode. Thus in the best case, compensation for an energy deficit of 2 GWh (for example) would require a minimum of eight 250 MW plants—assuming we could ramp these up to full capacity in minutes—which is not possible. Thus, more fast-response plants (possibly peakers) must be held in standby. Their fast response capacity is on the order of 20 to 30 percent of nameplate capacity. Thus, in this simple example, we would need up

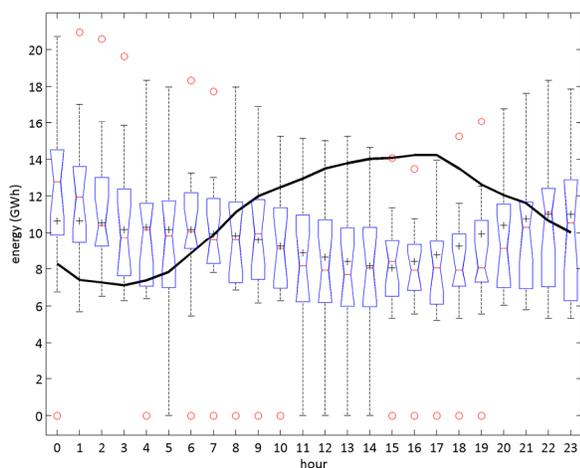

Fig. 7: Energy demand and production—the solid black line represents the demand required to meet 15% of energy needs. The boxplot captures the inherent variation in the energy generated from wind.

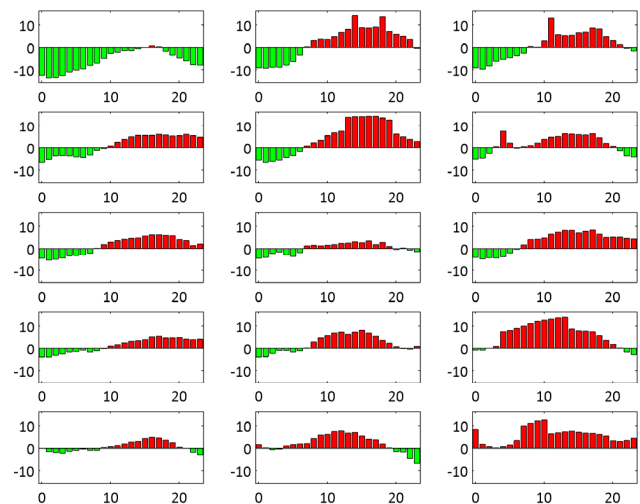

Fig. 8: Bar graphs showing hourly deficit (red) and surplus (green) energy produced if wRPS15 is achieved. The ordinate units are GWh. Surplus amounts are generated during early morning hours when demand is lowest.

to 40 such generation plants.

4. What is the critical penetration threshold beyond which wind energy begins to pose systemic deficit risk to California's grid? As noted earlier, California's energy grid has between 5 and 10% (about 2.5 to 5 GWh) reserve capacity. As shown in Fig. 9, the maximum energy deficit outliers of 14.23 GWh occurs at 1600h and 1700h. With appropriate scaling, the critical threshold is about 5% RPS, given a 5GWh reserve capacity (see Fig. 10). Beyond this threshold, compensating for wRPS energy deficits may consume all the reserve capacity—potentially leaving no capacity for other temporary generation shortfalls. Case in point: in 2008, Texas, the state with the largest fraction of wind in its energy portfolio (ca. 3% penetration level), suffered unexpected blackouts because wind speeds dropped below usable limits as a sudden gust of cold weather engulfed the state [19]. The experience of the Texas grid is another illustration of the systemic risks posed by increasing the fraction of wind beyond the energy reserves in the grid.

5. While we emphasize energy deficits in this work, wind intermittencies pose a risk when the deficits or surplus energy are too large. System operators are required to implement "load shedding" if the imbalance between generation and load demand is too large.

6. Hour-to-hour compensation requires additional wind forecasting tools beyond the methods described in Fig. 7 and Fig. 9. The hour-to-hour auto-correlation function (hhACF) shown in Fig. 11 illustrates the probabilistic reliability of wind energy forecasts for the 15 wind energy generation profiles in Fig. 8. We use the boxplot [15] to show the probability distribution for various hour-lags. Of interest are the 1-hour, 2-hour and 3-hour lags. The highest predictive certainty is for 1-hr lags. In this example, the worst case 1-hr compensation estimate in Fig. 9 has an uncertainty of 28.42%.[6] hhACF lags are a powerful wRPS compensation tool because they allow us to determine how much worst-case capacity must be online within a three hour window. This tool allows us to configure the optimal number of compensation plants for each lag window. The fastest compensation plants, e.g., combined cycle gas fired, may constitute the bulk of 1-hr lag compensators. However, 2-hr lag compensators may utilize other slower plants. 3-hr lag compensators may provide incremental base-load capacity and may indeed be coal-fired.

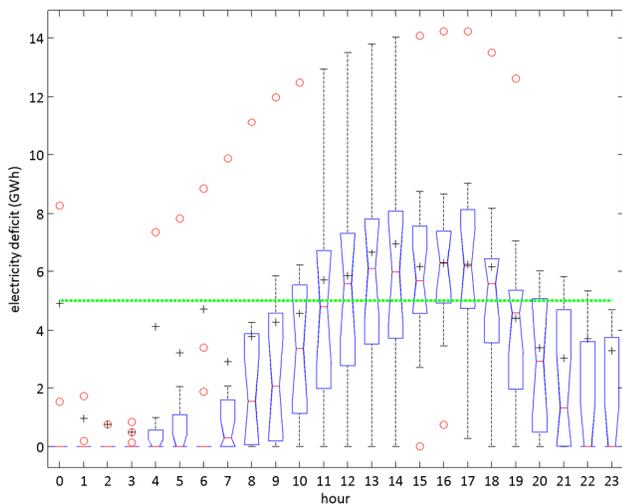

Fig. 9: Boxplot of the energy deficits from Fig. 8 vs. hour of day. The deficit variability reflects the unpredictability of wind power—even in the early morning hours when demand is lowest, significant deficits can result from intermittency.

[6] Note: 28.42% uncertainty is the worst case for a 95% confidence limit of the 1-hr lag. Fig. 11 actually shows one outlier corresponding to an uncertainty of 61.55%. The choice is a system design parameter.

7. What are the $CO_2$ and GHG costs associated with maintaining a large number of fossil-based generation plants in standby? While we do not put a price on carbon today, the options do not look good for wRPS15. Basically, each wRPS plant requires an array of fossil-based energy generation plants. So the macro exchange cost associated with replacing fossil-based generation capacity with wind energy is much lower than expected. From our estimates, the state would only reduce emissions by a small fraction. In macro exchange terms, if the state deploys 15% energy from wind, the expected payoff should be 15% reduction in $CO_2$ and GHG. With the compensation generation required to make wind energy stable, the actual reduction in $CO_2$ may end up fewer than 5% in the wRPS15 scenario. The problem is made worse because a FFP in standby mode is much less efficient than in energy production mode.

8. With a much lower than expected $CO_2$ and GHG payoff, the state of California must reevaluate the opportunity cost implications again. Beyond a certain threshold, it is not clear that wRPS is the best choice. It is possible that the opportunity costs of wRPS15 may be too large. For example, the capital costs of renewables and associated fast response generators must be revaluated against other choices. It is possible that the infrastructure to make wRPS15 work may end up more expensive in terms of capital dollars and its inability to curtail $CO_2$ and other GHGs.

We have, to this point, not factored the serious constraints posed by the grid infrastructure. In general, the most geographically optimal wRPS locations may not already host

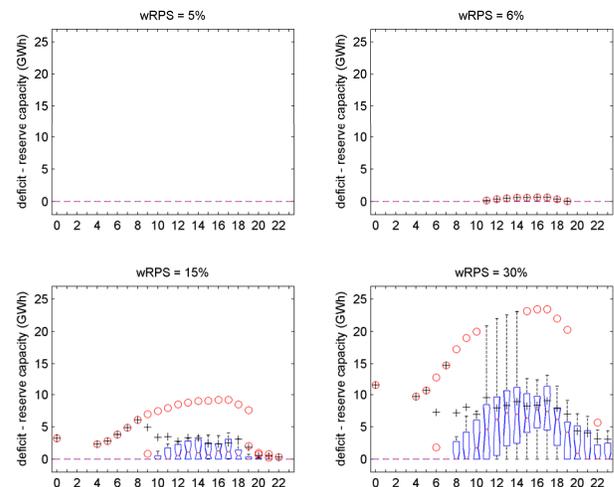

Fig. 10: Distribution of deficits exceeding reserve capacity of 5 GWh; increasing wRPS levels result in significant deficits above ~5%. The abscissae indicate the hours of day.

FFPs. Thus, the state needs to reconfigure the electric grid to make it possible to achieve wRPS15 and its associated deficit compensation FFPs.

### F. wRPS Risk Profile: Probability of multi-hour deficits

Consistent with the definition of risk in Section II-A, this section presents a brief illustration of the probabilities of energy deficits. From the scenario of 15 contiguous days in Fig. 2, our model calculates the probability of contiguous energy deficit clusters ranging from 1 to 15 hours for several wRPS penetration levels and reserve capacities. Energy deficit is defined as the difference between the hourly demand and the hourly RPS generation plus some reserve capacity. Fig. 12 shows a sample risk profile for wRPS = 15% and reserve capacity of 5 GWh. This model assumes the best-case scenario that all the reserve capacity can be used as the first-line compensation for wRPS energy deficits. The plot shows the probability of one and multiple-contiguous hour clusters of deficits. So, for example, 2 on the x-axis is the probability of a deficit lasting two hours. In most cases, the calculation metric shows a probability range, as opposed to a single probability. This formulation is necessary to illustrate the probability of a specific $n$-hour cluster without loss of accuracy.

From Fig. 12, we observe that the probabilities of deficits are large. The many serious impacts associated with these short-term intermittencies in wind energy are discussed in the follow-on paper on risk.

### G. Benefits of hour-by-hour correlation in wind speed

There is an intuitive conjecture that suggests that more wRPS dispersed over large geographies makes the energy profile of wind more stable from hour to hour. This is technically true as borne by marginal analysis in this work. However, our analysis shows two issues. True, more dispersion of wind farms reduces the probability of total energy outage—this is true because the wind sources would be less correlated. However, the aggregate wRPS energy from such a dispersed system is potentially much lower. The size of dispersed wind farms may also be much smaller than if they are located in wind-optimal geographical locations. In the best case, production of maximum energy from all available wind farms requires very high hour-to-hour correlation in wind speeds. This suggests that maximum energy output is inversely related to the reliability garnered from a widely dispersed wind farm infrastructure.

## III. Solutions

If current growth projections hold, wind energy will constitute 15% of California's energy portfolio by 2016; i.e., wRPS15 by 2016. However, the hour to hour intermittencies inherent in wind constitute significant and expensive systemic risks that are not adequately factored in our current energy policy. We agree that wind energy should indeed be part of California's RPS, but if the state is to satisfy 100% demand 100% of the time, wind energy requires radically different thinking. We suggest the following cost-contained solutions:

1. Throttle back wRPS15 to less than wRPS7.5 because the former increases the systemic risk of energy deficits significantly. Since the implementation costs of associated fast compensation plants are high, it is questionable whether achieving wRPS15 is the right choice for California given serious opportunity cost implications. From Fig. 12, the probability of multi-hour wRPS deficits are large at wRPS15—even with a hypothetical best-case application of 5 GWh "fast-response" reserve capacity.

2. Change the policy formulation toolset to that suggested in this paper. The toolsets used to plan for RPS energy generation need to include robust hour-to-hour analytics. As shown in our Fig. 7 and Fig. 8, the RPS decisions need to be built on hour-to-hour metrics that do not use monthly (or annual) averages. Current macro methods mask sub 1-hr intermittencies in wind energy and other RPS components.

3. California requires three classes of fast response (possibly peaker) generation plans consistent with the 1-hr, 2-hr and 3-hr lag described in bullet 4 of the Implications section. Careful analysis using the hhACF is essential to limiting the implementation cost of wRPS15 compensation.

4. Geographic location of wind farms must be carefully

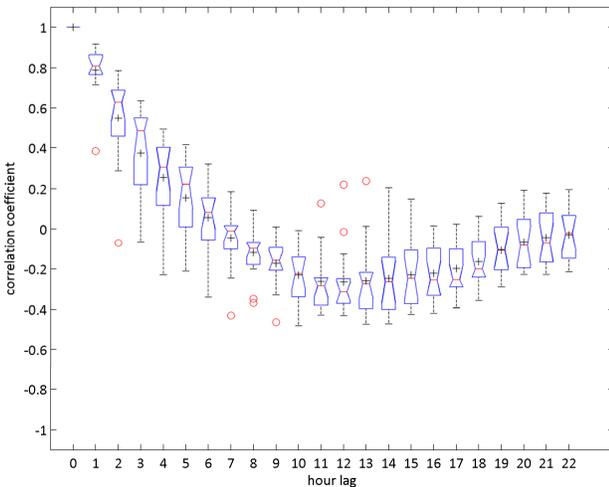

Fig. 11: Hour-to-hour autocorrelation coefficients of wind speeds vs. hour-lags. Low correlation coefficients prevent prediction of wind speeds (and energy) more than an hour or two in advance.

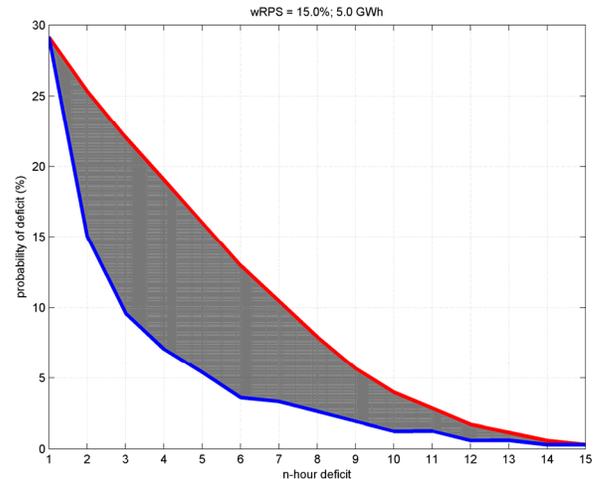

Fig. 12: Probability of $n$-hour energy deficit clusters for wRPS = 15% and reserve capacity = 5 GWh. The red and blue lines bracket the range of probabilities calculated two different ways.

balanced with energy stability needs. As shown in this, and future, work, maximum energy generation is somewhat at odds with wRPS15 system reliability.

5. Integrate wind farms into combined-cycle generation plants; i.e., as opposed to building a 500 MWh isolated wind farm, there is larger payoff with building a smaller wind farm that can integrate with other elements in a combined-cycle plant. As we have shown above, the probability variability of wind may reduce such a 500 MWh plant to less than 50 MWh on an hour-to-hour basis. Such close integration with other combined-cycle elements allows for easier compensation.

6. The state should avoid investing in large scale projects focused on bringing transmission lines to wind rich geographies if there are no other compensation plants nearby.

7. Wind energy mitigation strategy necessitates investment in utility-scale storage infrastructure. Indeed, storage is the underlying enabler for wind, solar and tidal components of the RPS.

While RPS thinking suggests that California may wean itself from importing nearly 25% of its energy, implementing RPS33 requires the state to integrate more with neighboring states with similar or different energy portfolios. Similar to Denmark, tighter integration with large hydro or nuclear from neighboring states provides necessary compensation for intermittencies in wind and other RPS components.

## V. BIOGRAPHIES

**Sam O. George** (BS 1993) is an EE graduate of Iowa State University, Ames, Iowa.

His competencies span a range of practices—high frequency / fidelity IC SoC product design / development (high frequency communications chipsets, power control subsystems and chipsets, multi-loop timing recovery designs, e.g., Fractional-N PLLs, data converters, et. al), BIST, systems and software architecture / modeling, SaaS, computer algorithms, Intellectual Property monetization and business problem solving / BPO.

Sam has led GridByte® for the past seven years, a multi-practice consultancy that, among other things, produces a range of risk-optimized strategic decision analytics tools / solutions for corporate and government clients. This innovation to management consulting combines engineering science and applied mathematics with systems expertise.

In addition to consulting positions at a number of fortune 500 companies, his career experience includes design management and IC lead positions at GlobespanVirata and Hughes Network Systems. In these positions he led high-frequency CMOS / BiCMOS chipset designs and had direct design responsibility for RF / mixed-signal / analog sub-systems, systems architecture, device modeling, yield optimization and DFM.

**H. Bola George** (Ph.D. 2007) is an applied physics graduate of Harvard University.

At GridByte, Inc., Bola has been focused on applying analytical frameworks to modeling of real-time physical systems, in particular, energy. Prior to employment at GridByte, Bola's work spanned investigation of surface mass transport mechanisms governing the formation of nanoscale features, materials development, and involvement with development of analytical tools to quantify morphological features.

**Scott V. Nguyen** (Ph.D. 2006) is a Senior Physicist in Shell's Innovation and R&D division where he identifies and develops technology applicable to the sustainable development of unconventional hydrocarbon resources, focusing on techno-economics, energy management, and greenhouse issues. He is a physics graduate of Harvard University and also currently serves on the advisory committee to the American Institute of Physics Corporate Associates.